\documentclass[
superscriptaddress,
twocolumn,
prab,
]{revtex4-1}

\usepackage{graphicx}
\usepackage{dcolumn}
\usepackage{bm}
\usepackage{amsmath,amsthm,amssymb}
\usepackage[separate-uncertainty = true]{siunitx}
\usepackage{xcolor}
\usepackage[english]{babel}
\usepackage{amsfonts}
\usepackage{soul}

\begin{document}
\preprint{APS/123-QED}

\title{Demonstration of kilohertz operation of Hydrodynamic Optical-Field-Ionized Plasma Channels}

\author{A. Alejo}\email{aaron.alejo@usc.es}
\affiliation{John Adams Institute for Accelerator Science and Department of Physics,University of Oxford, Denys Wilkinson Building, Keble Road, Oxford OX1 3RH, United Kingdom}%
\affiliation{Instituto Galego de F\'isica de Altas Enerx\'ias, Universidade de Santiago de Compostela, Santiago de Compostela 15782, Spain}%
\author{J. Cowley}%
\affiliation{John Adams Institute for Accelerator Science and Department of Physics,University of Oxford, Denys Wilkinson Building, Keble Road, Oxford OX1 3RH, United Kingdom}%
\author{A. Picksley}%
\affiliation{John Adams Institute for Accelerator Science and Department of Physics,University of Oxford, Denys Wilkinson Building, Keble Road, Oxford OX1 3RH, United Kingdom}%
\author{R. Walczak}%
\affiliation{John Adams Institute for Accelerator Science and Department of Physics,University of Oxford, Denys Wilkinson Building, Keble Road, Oxford OX1 3RH, United Kingdom}%
\author{S. M. Hooker} 
\affiliation{John Adams Institute for Accelerator Science and Department of Physics,University of Oxford, Denys Wilkinson Building, Keble Road, Oxford OX1 3RH, United Kingdom}%

\date{\today}

\begin{abstract}
We demonstrate experimentally that hydrodynamic optical-field-ionized (HOFI) plasma channels can be generated at kHz-scale pulse repetition rates, in a static gas cell and for an extended period. Using a pump-probe arrangement, we show via transverse interferometry that  the properties of two HOFI channels generated \SI{1}{ms} apart are essentially the same. We demonstrate that HOFI channels can be generated at a mean repetition rate of \SI{0.4}{kHz} for a period of 6.5 hours without degradation of the channel properties, and we determine the fluctuations in the key optical parameters of the channels in this period. Our results suggest that HOFI and conditioned HOFI channels are well suited for future high-repetition rate, multi-GeV plasma accelerator stages.
\end{abstract}

\maketitle

\setcounter{page}{1}
\section{Introduction}
Laser wakefield electron acceleration has attracted significant attention over the last decades, due to its ability to generate acceleration gradients three orders of magnitude greater than those achieved in conventional accelerators \cite{Tajima1979, Esarey2009, Hooker2013jk}. Currently, laser-wakefield accelerators can routinely generate stable electron bunches with ultra-short duration and GeV-scale energies \cite{Leemans:2006, Leemans:2014kp, Wang:2013el, Gonsalves:2019ht}, from acceleration stages a few centimetres long.

In a laser wakefield accelerator (LWFA), a laser pulse propagates through a plasma, which partially separates the plasma electrons and ions to drive a plasma (or `Langmuir') wave. The intense electric field within this wave can be of order $\SI{100}{GeV.m^{-1}}$, and this can be used to accelerate charged particles to high energies in a short distance. In order to excite a large amplitude plasma wave: (i) the duration of the laser should be less than the plasma period $T_\mathrm{p} = 2 \pi /\omega_\mathrm{p}$, where $\omega_\mathrm{p} = (n_\mathrm{e} e^2 / m_\mathrm{e} \epsilon_0)^{1/2}$, and $n_\mathrm{e}$ is the plasma electron density; and (ii) the peak laser intensity should be of order $\SI{E18}{W.cm^{-2}}$.

The energy gain of a laser-plasma accelerator, and the required length to achieve this, scale approximately as $\Delta W \propto 1/n_\mathrm{e}$ and  $L_\mathrm{acc} \propto 1/n_\mathrm{e}^{3/2}$, respectively \cite{Esarey2009}. These scalings require multi-GeV plasma accelerators to operate at plasma densities of order $\SI{E17}{cm^{-3}}$, with lengths of tens of centimetres. Since $L_\mathrm{acc}$ is considerably longer than its Rayleigh range, the focused driving laser pulse must be guided over the length of the accelerator.

Many potential applications of plasma accelerators --- such as compact light sources \cite{Fuchs:2009,Kneip:2010, Corde:2011bc, Phuoc:2012vb, Powers:2013, Khrennikov:2015gx}, including free-electron lasers \cite{Wang:2021} , and future particle colliders \cite{ALEGRO:2019} --- require the generation of multi-GeV electron bunches at high ($\gtrsim \SI{1}{kHz}$) pulse repetition rates, and with high stability over extended periods. Hence a major goal of current research in the field is the development of high-intensity waveguides capable of meeting these challenging requirements. 

Several methods for optically guiding high-intensity pulses have been demonstrated experimentally. Laser pulses with a peak power above the critical power \cite{Sprangle:1990a} are focused by the relativistic response of the plasma, and so can overcome diffraction. Relativistic guiding of this type has been demonstrated \cite{Poder:2017gt} over distances of up to \SI{90}{mm}, and has been exploited to generate multi-GeV electron beams \cite{Kneip:2009, Wang:2013el, Kim2013}. However, stable relativistic guiding requires drive beams of high mode quality \cite{Wang:2013el}, and is sensitive to shot-to-shot variations of the wave front of the drive laser \cite{Poder:2017gt}. 

An alternative, potentially more stable, approach is to employ a plasma waveguide, i.e.\ a column of plasma in which the electron density increases --- and hence the refractive index decreases --- with radial distance from the axis. A wide variety of methods have been explored for creating plasma waveguides, including hydrodynamic plasma channels \cite{Durfee:1993, Volfbeyn:1999cj, Smartsev:2019cy},  Z-pinches \cite{Hosokai:2000, Luther:2005eu}, open-geometry discharges \cite{Lopes:2003}, ablated-capillary discharges \cite{Ehrlich:1996}, gas-filled capillary discharges \cite{Spence:2000fr, Butler:2002zza, Gonsalves:2016jc}, and laser-heated gas-filled capillary discharges \cite{Bobrova:2013jz, Pieronek:2020bo, Gonsalves:2020bg}. Of these, the one used most successfully for laser-driven acceleration is the gas-filled capillary discharge waveguide and its laser-heated variant. For example, electron acceleration to an energy of \SI{7.8}{GeV} was recently demonstrated in a laser-heated capillary discharge waveguide \cite{Gonsalves:2019ht}. Although these results are impressive, this type of waveguide is prone to laser damage, which is likely to make it difficult to achieve operation at high repetition rates for extended periods.

\begin{figure*}[tb!]
    \centering
    \includegraphics[width=\textwidth]{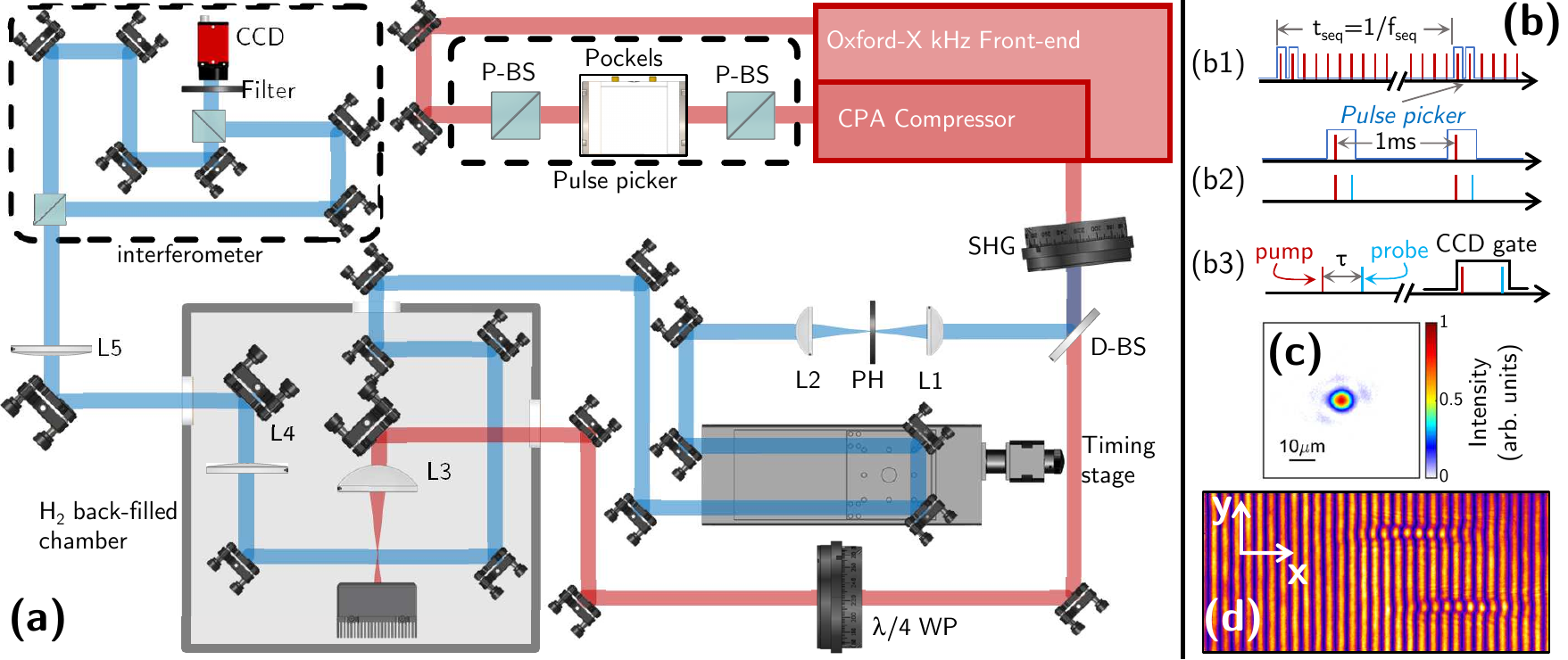}
    \caption{\textbf{(a)} Schematic diagram of the experimental setup used to generate HOFI plasma channels at kHz repetition rates. See the main text for a detailed description. \textbf{(b)} Schematic diagram of the generation of probe and pump pulses, showing: (b1) the train of \SI{800}{nm} pulses from the laser front-end, and the selection of pairs of \SI{800}{nm} pulses, the pairs being separated by $t_\mathrm{seq} = 1/f_\mathrm{seq}$; (b2) the conversion of each pair of pulses into a pump-probe pair; and (b3) gating of the CCD to record the interferogram generated by the second pump-probe pair. \textbf{(c)} Measured focal spot of the channel-forming beam. \textbf{(d)} Raw interferogram measured \SI{50}{ps} after the passing of the channel-forming beam, using a \SI{350}{mbar} backfill pressure of H$_2$.} 
    \label{fig:setup}
\end{figure*}

We have recently investigated a new type of waveguide which could meet the challenging requirements of multi-GeV, high-repetition-rate plasma accelerators. In hydrodynamic optical-field-ionized (HOFI) plasma channels \cite{Shalloo:2018fy, Shalloo:2019hv}, a column of plasma is formed and heated by optical field ionization (OFI). This column expands radially into the surrounding cold gas, driving a cylindrical shock wave, within which a plasma channel is formed. This concept is closely related to earlier work on collisionally-heated hydrodynamic channels \cite{Durfee:1993, Durfee:1995gr}, but using OFI, rather than laser-driven collisions, allows the formation of plasma channels at the lower densities needed for multi-GeV acceleration. Guiding of high-intensity pulses in HOFI channels as long as \SI{100}{mm} has been recently demonstrated \cite{Picksley:2020ec}, with on-axis densities below \SI{1e17}{\per\cubic\cm}. We have also investigated a variant of the HOFI scheme, conditioned HOFI (CHOFI) channels \cite{Picksley:2020ik}, in which the transverse wings of an intense laser pulse guided by the HOFI channel ionizes the collar of neutral gas surrounding the shock front, to form a deep, low-loss plasma channel. Related work on enhanced guiding via ionization of the neutral gas collar by a low-order Gaussian pulse \cite{Feder:2020gt} or by a high-order Bessel beam \cite{Miao:2020ir} has also recently been reported.

The parameters of (C)HOFI channels makes them very well suited as multi-GeV plasma accelerator stages. In addition, and very importantly, since they are free-standing they should be immune to laser damage, and hence they have the potential to be operated at high pulse repetition rates for extended periods. To date, however, this potential has not been explored experimentally, and the highest repetition rate at which HOFI channels have been generated is \SI{5}{Hz} \cite{Shalloo:2019hv}. Indeed, several factors could limit the maximum repetition rate at which these HOFI channels could be operated. First, the plasma must not only recombine to form a neutral gas, but it must also return to a uniform density. Secondly, the operation at high repetition rates involves the deposition of a significant amount of energy that could result in an overall heating that could modify the plasma evolution, and therefore the channel generation. Finally, potential applications of these channels in guiding and acceleration require the channel parameters to remain stable, and therefore their robustness against jitter of the laser, alignment and gas conditions needs to be shown.

In the present paper, we demonstrate that HOFI plasma channels can be generated at repetition rates of at least \SI{1}{kHz} in a statically-filled gas cell, by showing that the temporal evolution of the transverse electron density profiles of the HOFI channels produced $\SI{1}{ms}$ apart are essentially the same. Furthermore, we demonstrate that the plasma channels generated at a mean pulse repetition rate of $\langle f_\mathrm{rep} \rangle = \SI{0.4}{kHz}$ are stable over a period greater than 6 hours, corresponding to more than 4 million shots.

\section{Experimental setup}
The experiments were carried using the kHz repetition-rate front-end of a \SI{12}{TW} Ti:sapphire laser system. A schematic of the experimental setup is shown in Fig.~\ref{fig:setup}. The front-end of the laser system delivered an energy of \SI{1.05(2)}{mJ} per pulse, compressed to a pulse duration of \SI{39(2)}{fs} full-width-at-half-maximum (FWHM) and collimated to a beam diameter of $\varnothing \simeq \SI{5}{mm}$. In order to investigate the formation of channels at different repetition rates, a pulse picker --- comprising a pair of polarising beam splitters (P-BS) and a Pockels cell --- was located before the CPA compressor. The pulse picker could select two or more pulses spaced by an integer multiple of \SI{1}{ms}.

The experiments employed a pump-probe arrangement in which the pump (or `channel-forming') pulse generated the HOFI plasma channel, and the probe pulse was used to characterise its evolution. In order to generate the pump and probe beams, pulses selected by the pulse picker, and compressed by the grating compressor, were propagated through a Second Harmonic Generating (SHG) crystal. For each pulse incident on the SHG crystal, a small fraction was converted to a probe pulse with a wavelength of \SI{400}{nm}. The remainder of the incident pulse, with a wavelength of \SI{800}{nm} and an energy of $\sim$\SI{1}{mJ}, constituted the co-propagating channel-forming pulse. The channel-forming and probe pulses were separated by a dichroic beamsplitter (D-BS), which reflected the \SI{400}{nm} probe beam and transmitted the \SI{800}{nm} driver beam.

The \SI{800}{nm} channel-forming beam was passed through a quarter-wave plate (WP) to convert its polarisation to circular, directed into a statically-filled vacuum chamber through an AR-coated window, and focussed with an $f'=\SI{30}{mm}$ convex lens (L3) to a focus of \SI{5}{\micro\metre} FWHM. Plasma channels were formed by filling the chamber with H$_2$ gas at a pressure in the range  $P_\mathrm{b}=$\SI{50}{mbar} to \SI{500}{mbar}. A beam dump was placed behind the focus to absorb the laser energy remaining after the generation of the plasma channel.

The \SI{400}{nm}-probe beam was spatially filtered by a pair of lenses (L1,L2) and a \SI{5}{\micro\m} pinhole (PH) in order to improve the spatial quality and contrast of the interferogram. The beam was then directed to a retro-reflecting timing stage, and  into the back-filled chamber so as to cross, perpendicularly, the focus of the channel-forming beam. The timing stage allowed the delay $\tau$ between the arrival of the channel-forming pulse and the arrival of the probe pulse (at their intersection) to be varied in the range $\tau=$\SI{-1}{ns} to \SI{4}{ns}. 

The probe pulse was coupled into, and out of, the chamber via optically flat BK7 windows. After leaving the chamber, the probe was  imaged onto a 12-bit, 5 megapixel CMOS camera by a Keplerian telescope of magnification $M=8.4$ comprising a pair of lenses (L4, L5). A folded-wave Mach-Zehnder interferometer was placed between the telescope and the camera. Several reference interferograms were recorded before and after plasma was generated by the channel-forming beam. The phase was retrieved from the recorded interferograms using a Fourier transform method \cite{Takeda:1982}. To retrieve the density profiles, a polynomial background subtraction was performed and the phase maps were Abel inverted. 

The maximum repetition rate at which data could be recorded was limited by the readout time of the camera, which was several milliseconds when the full field of view of the camera was recorded ($\SI{500}{\micro\metre}\times\SI{375}{\micro\metre}$). In order to prevent the camera from capturing  interferograms generated by multiple probe pulses, the pulse picker was used to produce either a single \SI{800}{nm} pulse, or a pair of such pulses separated by \SI{1}{ms}. For the case of two channel-forming pulses, the CMOS camera could be triggered to capture the interferogram generated by the second probe pulse only. By examining the channels formed by a single channel-forming pulse, and that produced by the second of a pair of such pulses, it was possible to investigate whether the plasma formed by the first pulse influenced the channel formed \SI{1}{ms} later. These pulse sequences were generated at a repetition rate of $f_\mathrm{seq} = \SI{10}{Hz}$, which was the rate at which data was collected.

\section{Experimental results}

\begin{figure}[tb]
    \centering
    \includegraphics[width=\linewidth]{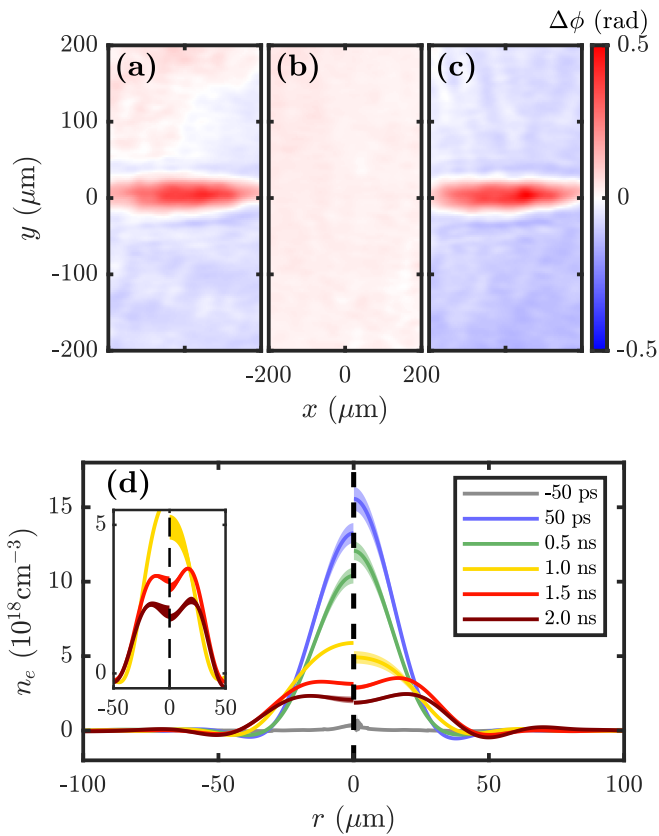}
    \caption{\textbf{(a,b,c)} Phase maps retrieved from the interferometric measurements. (a,c) show the plasma generated by (a) a single channel-forming pulse, and (c) the second of a pair of channel-forming pulses separated by 1 ms, measured at $\tau=\SI{0.5}{ns}$ after the passing of the channel-forming beam. (b) Measured phase shift $\SI{50}{ps}$ prior to the arrival of the second channel-forming pulse. \textbf{(d)} Temporal evolution of the transverse electron density profile $n_\mathrm{e}(r)$ produced by: (left) a single channel-forming pulse; and (right) the second of a pair of channel-forming pulses separated by \SI{1}{ms}. For each plot the line and shading show, respectively, the average and standard deviation of the density measured over 50 shots. For these data $P_\mathrm{b} = \SI{350(3)}{mbar}$. The inset in (d) depicts a zoom-in of the electron density profiles for $\tau\geq\SI{1}{ns}$, clearly showing the plasma channel formed at those times.}
    \label{fig:profiles}
\end{figure} 

\begin{figure*}[t!]
    \centering
    \includegraphics[width=\textwidth]{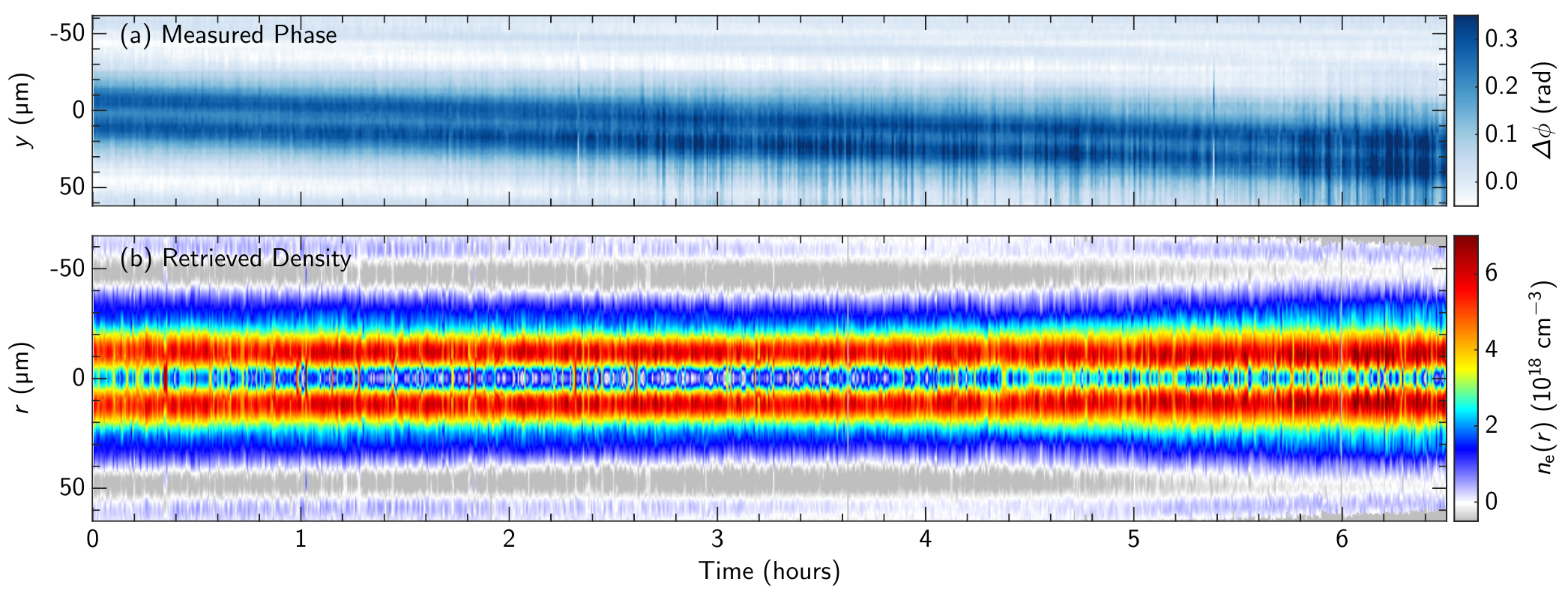}
    \caption{Demonstration of the long-term stability of HOFI channels. \textbf{(a)} The measured phase map; the apparent slow drift in the transverse position of the plasma is an artifact caused by a slow drift in the position of the probe beam. \textbf{(b)} The transverse electron density profiles of the HOFI channels formed at a delay $\tau = \SI{1.5}{ns}$ after the second of the pair of channel-forming pulses separated by \SI{1}{ms} are shown for data recorded at $f_\mathrm{seq} = \SI{200}{Hz}$ over a period of 6.5 hours. For these data the initial gas pressure was $P_\mathrm{b} = \SI{350}{mbar}.$}
    \label{fig:stability}
\end{figure*}

Figure \ref{fig:profiles} shows the temporal evolution of the transverse electron density profile $n_\mathrm{e}(r)$ of HOFI plasma channels generated by single channel-forming pulses, and by the second of a pair of such pulses separated by \SI{1}{ms}. The evolution of the channels formed by a single pulse is consistent with that we have reported previously \cite{Shalloo:2018fy, Shalloo:2019hv}. Initially ($\tau\simeq \SI{0}{ns}$), a cylindrical plasma column is generated with a diameter of $\sim\SI{25}{\micro\meter}$, and a peak electron density of $n_\mathrm{e}(0) \approx \SI{13e18}{\per\cm\cubed}$. This value is consistent with full ionisation of the ambient gas at $P_\mathrm{b} = \SI{350(3)}{mbar}$, which corresponds to an electron density of $\simeq \SI{17(2)e18}{\per\cm\cubed}$. The plasma column then expands rapidly, driving a cylindrical shock wave into the surrounding gas, and forming a plasma channel within the region defined by the shock front. The formation of a HOFI channel is evident for $\tau\geq\SI{1.5}{ns}$.

As seen in Fig.\ \ref{fig:profiles}, the channel formed by the \emph{second} of a pair of channel-forming pulses separated by \SI{1}{ms} is almost identical to that formed by a single pulse. It can therefore be concluded that, even with a static gas fill, the plasma cools and recombines to form an approximately uniform gas within \SI{1}{ms}, consistent with the absence of plasma in the interferogram measured immediately prior to the arrival of the second pulse.

The results of Fig.\ \ref{fig:profiles} demonstrate that the gas recovers to a state suitable for channel formation within \SI{1}{ms}, which suggests that operation at a repetition rate of at least \SI{1}{kHz} could be possible. However, that data was recorded at only $f_\mathrm{seq} = \SI{10}{Hz}$, and so any effects of long-term heating of the gas would not have been detected. In order to investigate this further, interferograms were recorded with a reduced field of view of $\SI{135}{\micro\metre}\times\SI{100}{\micro\metre}$, which allowed data to be recorded for pairs of channel-forming pulses separated by \SI{1}{ms} at a pulse sequence repetition rate of $f_\mathrm{seq} = \SI{200}{Hz}$, and hence a mean pulse repetition rate of $\langle f_\mathrm{rep} \rangle = \SI{0.4}{kHz}$.

Figure \ref{fig:stability} shows the transverse electron density profiles of the HOFI channels, formed at a delay $\tau = \SI{1.5}{ns}$ after the second of the pair of channel-forming pulses, recorded at $f_\mathrm{seq} = \SI{200}{Hz}$ over a period of 6.5 hours. It can be seen that HOFI channels were created throughout this period, with no significant long-term evolution of their properties. These data demonstrate that slow heating of the gas will not have a deleterious effect on the channels for mean repetition rates at least up to \SI{0.4}{kHz}.

\begin{figure}[tb]
    \centering
    \includegraphics[width=\linewidth]{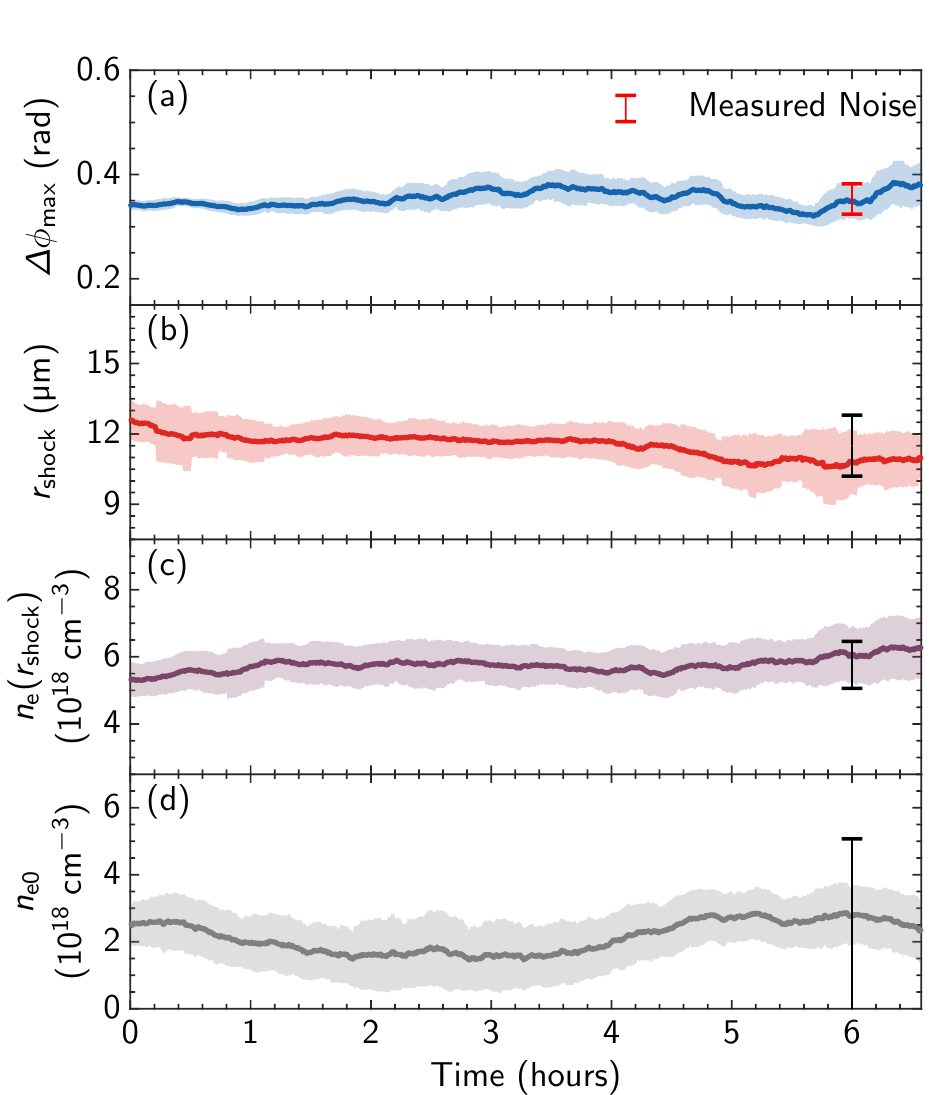}
    \caption{Long-term evolution of the total measured phase shift and the key optical properties of the $\langle f_\mathrm{rep}\rangle = \SI{0.4}{kHz}$ HOFI plasma channels shown in Fig.\ \ref{fig:stability}. In all cases, the solid lines and shading show, respectively, the moving average and standard deviation over 50 shots. \textbf{(a)} Maximum phase shift measured ($\Delta \phi_\mathrm{max}$). The red error bar indicates the standard deviation of the averaged measured noise for the shots in the run. \textbf{(b)} The measured shock position $r_\mathrm{shock}$, defined by the position of the peak electron density, \textbf{(c)} the electron density at the shock front $n_\mathrm{e}(r_\mathrm{shock})$, \textbf{(d)} the axial density $n_\mathrm{e0}$. For (b-d), the black error bars indicate the standard deviation of the plotted parameter deduced from a Monte Carlo simulation of the effect of noise on the Abel inversion. The $y$-position of the centre of the error bars is equal to the mean value during the run. The error bars are shown at $t=\SI{6}{hours}$; this is an arbitrary choice, although it does allow for direct comparison with the data at the time when the noise of the phase maps is highest.}
    \label{fig:properties}
\end{figure}

Applications of high-repetition-rate HOFI and CHOFI plasma channels will require their optical properties to remain stable for long periods of operation. Figure \ref{fig:properties} shows the variation over 6.5 hours of the peak phase shift, and the key properties of the HOFI channels. The maximum measured phase shift (shown in Fig.\ \ref{fig:properties}(a)) was $\Delta \phi_\mathrm{max} = \SI{0.35(4)}{rad}$, where the shaded area represents the root-mean-square (rms) value. By taking a portion of the recovered phase map that did not contain plasma, the average noise was found to be approximately $\SI{0.033}{rad}$. The red error bar in Fig.\ \ref{fig:properties}(a) indicates this. The noise was found to increase slightly during the run as the probe beam pointing drifted, and hence the reference shots (which were recorded at the beginning of the run) became less representative of the background phase present on data shots. It can be seen that the variation of the maximum phase shift  was within the variation expected from the level of noise in the data. 

The mean values of the key channel parameters were found to be $r_\mathrm{shock} = \SI{11.5(10)}{\micro m}$, $n_\mathrm{e}(r = r_\mathrm{shock}) = \SI{5.7(7)e18}{cm^{-3}}$, and $n_\mathrm{e0} = \SI{2.10(120)e18}{cm^{-3}}$. 
It is worth noting that the energy, pointing, and spatial phase of the laser were not stabilized, and it is expected that the fluctuations in the channel properties evident in Fig.\ \ref{fig:stability} could be further reduced with suitable pulse stabilization. However, (C)HOFI channels are very robust with respect to variations of the laser and gas conditions. One can consider, for instance, the importance of stabilising the pulse energy. 
Since OFI shows a threshold behaviour, a small increase in the laser pulse energy would increase the radius out to which ionization occurred, but the initial electron energy in the ionized region would not be changed. Sedov-Taylor blast wave theory predicts that the radial position of the shock front scales as $r_s\propto E_\mathrm{OFI}^{1/4}$, where $E_\mathrm{OFI}$, is the energy per unit length of the initial plasma column. This slow dependence means that fluctuations in the laser parameters are not expected to cause significant jitter in the parameters of the generated plasma channel. Considering the robustness of the mechanism, it is useful to investigate the extent to which the observed variation in these parameters are caused by noise in the data, rather than resulting from jitter in the parameters themselves. This is likely to be a particular issue when retrieving $n_\mathrm{e0}$, since it is known Abel inversion can amplify small levels of noise to yield large errors near the axis. We investigated these effects via Monte Carlo simulations in which Gaussian noise, with a mean and standard deviation equal to that measured in regions away from the plasma, was added to a representative, smoothed data phase map (see Supplemental Material \cite{supp_mat} for more detail). The black error bars in Fig.\ \ref{fig:properties}(b-d) show the calculated root-mean-square (rms) fluctuations in the retrieved parameters which were deduced from this study. It can be seen that for all of the retrieved channel properties, the observed variation is consistent with the fluctuations arising from noise in the measured phase maps. We conclude, therefore, that the jitter in the channel parameters observed in Fig.\ \ref{fig:properties}(b-d) is dominated by the effects of noise in the measurement and hence is \emph{not} representative of the true shot-to-shot variation in these parameters.

\section{Discussion and conclusion}

The plasma channels generated in this study were limited to a length of only $\lesssim \SI{1}{mm}$ by the available laser energy. Nevertheless, the findings will be applicable to much longer HOFI channels  --- such as the \SI{100}{mm} long channels generated with an axicon lens, which were recently reported \cite{Picksley:2020ec} --- since the energy deposited per unit length of channel is determined only by the density and radius of the initial plasma column, and hence will be comparable.

We have recently demonstrated an extension to the HOFI channel scheme which produces very deep plasma channels with much lower transmission losses \cite{Picksley:2020ik}. In conditioned HOFI (CHOFI) channels a high-intensity conditioning pulse is guided by a HOFI channel. The transverse wings of the conditioning pulse ionize the neutral collar of gas surrounding the HOFI channel to form a deep, low-loss CHOFI plasma channel. Channels with axial electron densities of  $n_\mathrm{e0} \approx \SI{1E17}{\per\centi\meter\cubed}$ and a matched spot size of $w_\mathrm{m} \approx \SI{25}{\micro m}$ have been generated by this method with a power attenuation length of up to \SI{26(2)}{m}. We note that Milchberg \textit{et al.} have explored a similar scheme in which the collar of gas is ionized by a self-guided pulse \cite{Feder:2020gt}, as described above, or by a high-order Bessel beam \cite{Miao:2020ir}.

It is interesting, therefore, to consider the extent to which the results presented here for HOFI channels could be applied to CHOFI channels. In the latter, the conditioning pulse ionizes the collar of gas out to a radius of $r_\mathrm{CHOFI} \approx \SI{50}{\micro m}$, compared to the radius of the initial plasma columns generated in this work of $r_\mathrm{HOFI} \approx \SI{30}{\micro m}$ (see Fig.\ \ref{fig:profiles}). As such, the ratio of the required energy deposited per unit length to generate a CHOFI channel, to that needed for the HOFI channels studied here, is approximately $(r_\mathrm{CHOFI}/r_\mathrm{HOFI})^2 \approx 2.8$. This ratio is not large, and hence the energy deposited in the plasma per unit length of CHOFI channel will not be very different from that in this experiment. It is therefore expected that the conclusions of this work will also apply to CHOFI channels.

We note that additional energy is likely to be deposited in these channels by the \emph{guided} pulse. For example, in a LWFA the wakefield energy remaining in the plasma after particle acceleration could be significant, and the residual plasma waves will create a highly non-uniform density and temperature distribution. These non-uniformities need to dissipate, or be removed, before the accelerator can be operated again. 

One approach is to use one or more trailing, out-of-phase laser pulses to reduce the amplitude of the residual wakefield. In this case the wakefield energy is removed from the plasma by the frequency up-shifted photons of the trailing laser pulse(s), and, in principle can be recovered, increasing the efficiency of the accelerator. In a first step towards demonstration of energy recovery in a plasma accelerator, experiments showed that the wakefield amplitude can be reduced by $(44 \pm 8)\%$ by a single out-of-resonance laser pulse \cite{Cowley:2017cf}.

If the time for the gas to return to its initial state --- by either passive or active processes --- is too long, an alternative approach would be to provide fresh gas for each shot by flowing the gas transversely. We can establish a conservative estimate for the upper limit of $f_\mathrm{rep}$ enabled by this method by assuming that the gas needs to be moved by a distance of $\sim \SI{100}{\micro m}$, and that it can be flowed at a speed of $\sim \SI{100}{m.s^{-1}}$ (substantially less than the speed of sound in hydrogen gas $c_\mathrm{s} \approx \SI{1300}{m.s^{-1}}$). These values yield a gas clearance time of \SI{1}{\micro s}, suggesting that repetition rates as high as \SI{1}{MHz} could be achieved in this scenario.

Turner \textit{et al.} \cite{Turner2021} recently characterised the stability of Hydrogen-filled capillary discharge waveguides over 2000 consecutive shots over a period of 30 minutes, and showed that the fluctuations in the matched spot size of the channel were $\lesssim \SI{0.1}{\%}$.

Our Monte Carlo analysis of the effects of noise in the channel parameters retrieved by Abel inversion showed that the observed shot-to-shot fluctuations in these parameters was dominated by noise. As such it is not possible to determine the jitter in the channel parameters from the data presented here.  Future experiments implementing additional diagnostics, such as spectral variations of the channel-forming pulse or characterisation of guiding by the channel, would allow to reduce the uncertainty in the channel parameters.  We note that the high repetition rates which can be achieved with HOFI and CHOFI channels would make it straightforward to employ feedback techniques to ensure that the channels have high stability.

In conclusion, we have demonstrated that a HOFI channel can be generated in a static gas cell \SI{1}{ms} after a previously generated HOFI channel, with no change in the channel properties. Further, we showed that HOFI channels could be generated at a mean pulse repetition rate of $\langle f_\mathrm{rep}\rangle = \SI{0.4}{kHz}$ for a period of 6.5 hours without degradation of the channel properties due to the effects of heating or damage to the laser optics. The lower loss CHOFI plasma channels are expected to behave similarly.

HOFI and CHOFI channels can be generated at low electron densities ($n_\mathrm{e0} \sim \SI{1E17}{cm^{-3}}$), with long power attenuation lengths ($L_\mathrm{att} \gtrsim \SI{10}{m}$), with lengths of order \SI{1}{m}, and since they are free-standing they are immune to laser damage. We have previously shown that the total energy of the channel-forming pulse(s) required to generate HOFI and CHOFI channels is approximately \SI{1}{mJ} per millimeter of channel \cite{Shalloo:2019hv,Picksley:2020ik}. Hence, $\sim \SI{100}{mm}$ long HOFI and CHOFI plasma channels could be generated at high repetition rates with with state-of-the-art \SI{100}{mJ} class, \SI{1}{kHz} ultrafast lasers \cite{Budriunas:17}. The results presented here show that such channels could be generated for extended periods, which suggests that they would be ideal for future high-repetition rate, multi-GeV plasma accelerator stages.

\section{Acknowledgements}
This work was supported by the UK Science and Technology Facilities Council (STFC UK) [grant numbers ST/P002048/1, ST/R505006/1]; the Engineering and Physical Sciences Research Council [EP/V006797/1]. This material is based upon work supported by the Air Force Office of Scientific Research under award number FA9550-18-1-7005. This work was supported by the European Union's Horizon 2020 research and innovation programme under grant agreement No. 653782. This work was supported by `la Caixa' Foundation (ID 100010434) [fellowship code LCF/BQ/PI20/11760027]; Xunta de Galicia (Centro singular de investigaci\'on de Galicia accreditation 2019-2022); European Union ERDF; and the `Mar\'ia de Maeztu' Units of Excellence program MDM-2016-0692 and the Spanish Research State Agency.

Cite as \url{doi.org/10.1103/PhysRevAccelBeams.25.011301}. All the data used for the results can be downloaded from \cite{DataLink}.

\bibliography{kHz}

\end{document}